# The causal structure of Minkowski space time – possibilities and impossibilities of secure positioning

*Muhammad Nadeem*
*Department of Basic Sciences,*
*School of Electrical Engineering and Computer Science*
*National University of Sciences and Technology (NUST)*
*H-12 Islamabad, Pakistan*
*muhammad.nadeem@seecs.edu.pk*

Secure positioning, a prover located at a specified position convinces a set of verifiers at distant reference stations that he/she is indeed at the specific position, is considered to be impossible if the prover and verifiers have no pre-shared data while dishonest provers have an arbitrary amount of pre-shared entanglement [*Nature* 479, 307-308 (2011)]. We argue here that current impossibility results for secure positioning are the upshot of not utilizing full powers of relativistic quantum information theory and show that secure positioning and hence position-based quantum cryptography is possible if causal structure of Minkowski space time and quantum non-locality is used properly.

## 1. Introduction

Information theory deals with compressing, storing, processing and secure communication of data. It has deep connections with applied mathematics, computer science, electrical engineering, and physics while has important applications in number of other fields and multidisciplinary understanding. For example, its impact is crucial from mobile communication to internet, from neurobiology to space war, from thermal physics to understanding of black holes, and so on. The information theory has evolved the world progressively and securely through multiple ways such as source coding, channel coding, algorithmic information theory and information-theoretic security and measures of information.

In today's digital world, one of the most important areas in information theory is information-theoretic security such as *one-time* pad. Shannon's mathematical information theory [1] is based on deterministic systems (0 and 1) for encoding information where security relies on following three main cryptographic techniques: (i) symmetric encryption, (ii) asymmetric encryption, and (iii) hashing along with message authentication codes and digital signatures. However, widely used classical algorithms for distribution of symmetric keys, generation of public-private key pairs and hence hashing or digital signature are only computationally secure – eavesdroppers with efficient technology (quantum computer) can easily break all these mathematically hard algorithms.

On the other hand, newly developed Wiesner's quantum information theory [2] encodes information over probabilistic microscopic physical systems called qubits $|\varphi\rangle = \alpha|0\rangle + \beta|1\rangle$; an atom, nuclear spins, or polarized photon. These encoded quantum systems are represented by unit vectors in Hilbert space and are processed through unitary operators. This framework of quantum information theory allows defining cryptographic tasks such as QKD [3] with information-theoretic security guaranteed by laws/properties of quantum physics such as uncertainty principle and no-cloning [4]. Moreover, quantum non-local correlations have high-flying technological [5-8] and imperative impacts on foundation of quantum mechanics [9-11].

However, an important task of secure positioning [12-15] is still an enigma in both classical [16] and quantum information theory [15,

17]. In a general position-verification scheme, a prover located at a specified position convinces a set of verifiers at distant reference stations that he/she is indeed at the specific position. In a formal notion of position-verification, different verifiers send a secret message and a key to decrypt that message in pieces to the prover. That is, each verifier sends a bit of key such that all the key bits and the message arrive at the position of the prover concurrently. If the prover decrypts the message correctly and sends the result to all verifiers in time, position-verification scheme enables the verifiers to verify his position jointly. But if one or a set of dishonest provers, not at the specified position, intercept the communication and try to convince verifiers that they are at the specified position, a secure position-verification scheme enables the verifiers to reject it with high probability.

After a number of attempts made for achieving secure positioning, currently it is known that if the prover and verifiers do not pre-share any data between them and dishonest provers have an arbitrary amount of pre-shared entanglement then secure positioning of the prover is impossible [15-17]. After a series of attempts and no-go theorem for secure positioning, some authors showed that position-verification can be possible in the following two models: (i) bounded storage model where dishonest provers' quantum memory is bounded [15,18] or (ii) if the prover and verifiers have pre-shared data among them [19,20]. These result are useful somehow in the sense that position-verification gives a second layer of security, along with usual cryptographic techniques. However, we are not interested in such models here.

These impossibility results put a real qualm on the following fascinating thoughts: Can journey of information theory from Shannon's mathematical world "A mathematical theory of communication" to Wiesner's quantum world "conjugate coding" enter into a new regime where information-theoretic security would be based solely on positions? In other words, can we achieve information-theoretic security where only credential of communicating parties is their position? We show here that such unconditionally secure position-based quantum cryptography is possible where sender and receiver have no pre-shared quantum/classical data while dishonest provers have efficient quantum technology and unlimited computational powers.

Recently, we proposed that combination of quantum non-locality and causality allows relativistic quantum information theory to define a number of mistrustful cryptographic tasks [21-23]. For example, a new notion of oblivious transfer where both the data transferred and the transfer position remain oblivious, deterministic two-sided two-party secure computation, asynchronous ideal coin tossing with zero bias, and unconditionally secure bit commitment with arbitrarily long commitment time. These possibilities of wide range of cryptographic tasks motivate us to analyze also the possibilities of securing positioning in the same setup [24].

We argue here that existing no-go theorem for secure positioning are the upshot of not utilizing full powers of relativistic quantum information theory and show that secure positioning is possible if causal structure of Minkowski space time and quantum non-locality is utilized properly. Considering the possibilities of non-local instantaneous computations, we show that the possibilities or impossibilities of secure positioning really depend on the construction of position-verification scheme and its communication modes. First we analyze the impossibility for secure positioning on the basis of quantum communication modes (used previously for position-verification), geometric structure of Minkowski space time and causality. We move then to show possibility conditions for secure positioning while using full powers of quantum non-locality and causal structure of Minkowski space time. Finally, we propose a quantum scheme that guarantees unconditionally secure positioning and evades existing no-go theorem.

## 2. Causal structure of Minkowski space time and quantum non-locality

Einstein's relativistic laws, no superluminal signaling and covariance, can be interpreted in a purely geometric way in Minkowski space time where the curvatures are all zero and the geodesics are world-lines over specific hyper surface in four-dimensional coordinate system $(t,x,y,z)$. Before discussing possibilities and impossibilities of secure positioning, it is useful to revisit some of basic definitions and their terminologies in theory of relativity and quantum theory.

*Causality principle:* The causality principle does not allow specially separated observers P1 and P2 in Minkowski space time to communicate with one another instantaneously by sending superluminal signals. They can exchange their information only at space time position lying somewhere in causal future of both observers.

*Quantum non-locality:* Non-locality says that behavior of specially separated entangled quantum systems can depend on events occurring outside their respective light cones. It deviates from the understanding of locality where behavior of specially separated systems depends only on events occurring in their respective light cones.

*Localizable operators:* A bipartite superoperator $U = U_1 \otimes U_2$ is called localizable if it can be implemented by two distant parties $P_1$ and $P_2$ acting locally on their halves of shared quantum system $H_1 \otimes H_2$ such that $H_1 \otimes H_2$ transforms to $U(H_1 \otimes H_2)$ without any quantum/classical communication from $P_1$ to $P_2$ or from $P_2$ to $P_1$. In general, any tensor product $U = U_1 \otimes U_2$ is localizable if $P_1$ and $P_2$ have pre-shared entanglement between them. In short, if $P_1$ and $P_2$ are space-like separated, they can transform shared quantum system $H_1 \otimes H_2$ to $U(H_1 \otimes H_2)$ on same space-like hyper surface through localizable superoperator $U = U_1 \otimes U_2$.

*Causal operators:* A superoperator is said to be causal if it does not allow signaling between space-like separated parties. That is, a causal superoperator conveys information neither from $P_1$ to $P_2$ nor from $P_2$ to $P_1$. Hence, any localizable superopertor is a causal superopertor.

*Semi-localizable operators:* A bipartite superoperator $U = U_1 \otimes U_2$ is called semi-localizable if it can be implemented by two parties $P_1$ and $P_2$ acting locally on their halves of shared quantum system $H_1 \otimes H_2$ such that $H_1 \otimes H_2$ transforms to $U(H_1 \otimes H_2)$ with one-sided communication only; from $P_1$ to $P_2$ or from $P_2$ to $P_1$. For example, if actions of $P_1$ and $P_2$ are null-like separated with $P_2$ in causal future of $P_1$, they can transform shared quantum system $H_1 \otimes H_2$ to $U(H_1 \otimes H_2)$ through semi-localizable superoperator $U = U_1 \otimes U_2$ with one-sided communication from $P_1$ to $P_2$.

*Semi-causal operators:* A superoperator is said to be semi-causal if it does allow one-way signaling only. That is, a semi-causal superoperator can convey information from $P_1$ to $P_2$ (say) but not from $P_2$ to $P_1$. Hence, any semi-localizable superopertor is a semi-causal superopertor.

### 2.1 Quantum non-locality vs. causality principle:

From principles of these two different theories, someone may worry about following question: Are these principles of causality and quantum non-locality contradictory? It is really a demanding question as for as foundational perspectives are concerned; basic structures of theory of relativity and quantum theory are fundamentally different. Hence, comparison or combination of these two theories needs some serious attention. However, answer to the above question is No and we try to explain this in the information theoretic viewpoint; information is physical and can only be encoded over physical systems. Hence these information carrier physical systems must obey the laws of physics [25].

Suppose observers $P_1$ and $P_2$ have components, $H_1$ and $H_2$ respectively, of entangled quantum system $H = H_1 \otimes H_2$ at distant sites on some space-like hyper surface $S_T = \{t = T\}$. Quantum non-locality says that $P_1$ and $P_2$ can transform this unknown quantum system $H_1 \otimes H_2$ to $U(H_1 \otimes H_2)$ instantaneously through local unitary operations. Now if $U = U_1 \otimes U_1$ is a measurement operator, then respective measurement outcomes of $P_1$ and $P_2$ will be correlated. Both $P_1$ and $P_2$ surely can be aware of each other's measurement results but no information has been transferred from $P_1$ to $P_2$ or $P_2$ to $P_1$.

According to information theory, information can only be communicated between $P_1$ and $P_2$ at light speed through physical systems (photons) following world-lines over some null-like hyper surface $N_n$ connecting them. On the same footing, causality principle says that these world-lines must be directed in their future light cones. In short, after combining quantum non-locality, information theory and causality principle, we come to the following conclusion for non-local quantum measurement operation in Minkowski space time: specially separated $P_1$ and $P_2$ on hyper surface $S_{T_1} = \{t = T_1\}$ can send their measurement outcomes at light speed to p on space-like hyper surface $S_{T_2} = \{t = T_2\}$ or to each other only at points $q_1$ and $q_2$ lying on space-like hyper surface $S_{T_3} = \{t = T_3\}$ such that $T_3 > T_2 > T_1$. Hence non-local instantaneous measurements are not violating causality principle.

Detailed discussion on observables of relativistic quantum theory, possibilities and impossibilities of non-local instantaneous measurements, necessary and sufficient condition for no-communication theorem, and rigorous review on relativistic quantum information theory can be found somewhere else [26-30].

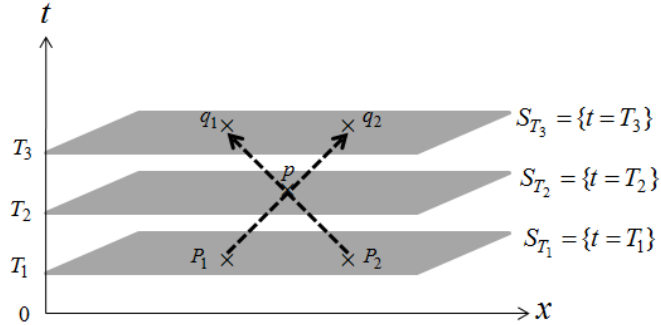

**Figure 1:** Specially separated $P_1$ and $P_2$ can perform localizable and hence causal operations on space-like hyper surface $S_{T_1} = \{t = T_1\}$. However, they can only perform semi-localizable and hence semi-causal operations with p, $q_1$ and $q_2$ only. In summary, $P_1$ and $P_2$ cannot send information to each other on same hyper surface $S_{T_1} = \{t = T_1\}$ but they can send information to p and $q_1$ or $q_1$ on hyper surfaces $S_{T_2} = \{t = T_2\}$ and $S_{T_3} = \{t = T_3\}$. However, p, $q_1$ or $q_2$ cannot send information to either $P_1$ or $P_2$.

## 3. Security model for secure positioning

For any general position-verification scheme, it is necessary to locate points from where verifiers send information to the prover, and where that information can be received and re-transmitted. For simplicity, let's consider a 1+1 dimensional case in Minkowski space time where prover P and verifiers $V_1$ and $V_2$ occupy sites at points $(x_p, 0)$, $(x_v^1, 0)$ and $(x_v^2, 0)$ respectively such that $x_p = (x_v^2 - x_v^1)/2$. Moreover, suppose there can be two dishonest provers $P_1$ and $P_2$ at points $(x_p^1, 0)$ and $(x_p^2, 0)$ respectively where $x_p^1 = x_p - \delta$ and $x_p^2 = x_p + \delta$ such that $\delta \ll x_p$. Now let's denote information sent from verifiers to the prover for position-verification as $I_p$ while information replied by the prover to the verifiers as $I'_p$.

We assume that all the verifiers and provers (honest or dishonest) lie in the same inertial frame, computation (local or non-local) time at their sites is negligibly small, and they can communicate quantum/classical signals with each other bounded by no-signaling principle. These assumptions are necessary to deal quantum information in relativistic setup where simultaneity, no signaling, and Lorentz invariance have fundamental importance.

In general, verifiers $V_1$ and $V_2$ encode information $I_p$ over composite quantum system $H = H_1 \otimes H_2$ where subsystem $H_1$ is kept by $V_1$ and $H_2$ by $V_2$. Here $H = H_1 \otimes H_2$ can be a product system, entangled system, or component of some larger quantum system $H' = H_1 \otimes H_2 \otimes H_V$. The prover (or dishonest provers) receives systems $H_1$ and $H_2$ from $V_1$ and $V_2$ respectively, applies some unitary transformations $U$ on $H = H_1 \otimes H_2$ depending on the scheme, and returns information $I'_p$ to both $V_1$ and $V_2$. The verifiers validate the exact position of the prover P if he replies correct information $I'_p$ consistent with $U(H)$ within allocated time.

If verifiers $V_1$ and $V_2$ send information $I_p$ encoded over system $H = H_1 \otimes H_2$ from points $(x_v^1, 0)$ and $(x_v^2, 0)$ respectively, the prover P can receive system $H = H_1 \otimes H_2$ at point $p(x_p, t_p)$ where $t_p = (x_p - x_v^1)/c$ while dishonest provers $P_1$ and $P_2$ can receive same system $H = H_1 \otimes H_2$ at points $p_1(x_p^1, t_p^1)$ and $p_2(x_p^2, t_p^2)$ respectively where $t_p^1 = t_p^2 = t_p - \delta$. If the scheme is secure, it enables the verifiers $V_1$ and $V_2$ to receive information $I'_p$ back at points $v_1(x_v^1, t_v^1)$ and $v_2(x_v^2, t_v^2)$ respectively such that $t_v^1 = t_v^2 = 2t_p$ with following guarantees: (i) only prover P at position to be verified $p(x_p, t_p)$ can send information $I'_p$ at both points $v_1(x_v^1, t_v^1)$ and $v_2(x_v^2, t_v^2)$. (ii) Irrespective of pre-shared entanglement and non-local computation resources, dishonest provers $P_1$ and $P_2$ at positions $p_1(x_p^1, t_p^1)$ and $p_2(x_p^2, t_p^2)$ respectively cannot send information $I'_p$ at both points $v_1(x_v^1, t_v^1)$ and $v_2(x_v^2, t_v^2)$.

**3.1. Localizable and causal computations:** The possibilities or impossibilities of security requirements as described above really depend on the construction of position-verification scheme and the resources dishonest provers $P_1$ and $P_2$ have. Suppose $P_1$ intercept quantum system $H_1$ from $V_1$ and $P_2$ receives quantum system $H_2$ from $V_2$. Moreover, suppose they also have arbitrary amount of pre-shared entanglement in the form of system $H_p = H_{p_1} \otimes H_{p_2}$. First of all, $P_1$ and $P_2$ need to extract information $I'_p$ from quantum system $H = H_1 \otimes H_2$ through non-local instantaneous computations. Secondly, whether P1 and P2 can respond valid information $I'_p$ to both verifiers in time or not depends on the construction of position-verification scheme and its communication modes.

Now the most relevant questions in the discussion of positioning at this stage is to ask (i) whether space-like separated $P_1$ and $P_2$ can have any *localizable* bipartite superoperator $U = U_1 \otimes U_2$ to transform system $H_{p_1} \otimes H_1 \otimes H_2 \otimes H_{p_2}$ to $U(H_1 \otimes H_2)$ ? (ii) If yes, whether that superoperator is causal or not. The answer to the first question is yes, shared entanglement allows $P_1$ and $P_2$ to act $U_1$ and $U_2$ locally on their halves without any communication between them. That is, they can transform $H_1 \otimes H_2$ to $U(H_1 \otimes H_2)$, without any classical communication, on the expense of their shared entanglement. This makes answer to the second question obvious; superoperator $U$ will surely be causal since any localizable superoperator is a causal superoperator. Hence $P_1$ and $P_2$ can instantaneously transform system $H_1 \otimes H_2$ to $U(H_1 \otimes H_2)$ but can only extract information $I'_p$ somewhere in their causal future by communicating their classical local measurement outcomes.

In conclusion, possibilities of having localizable operators in dishonest provers toolkit leads to impossibilities of hiding information $I'_p$ from them. Hence, possibilities or impossibilities of secure positioning really depend on the construction of position-verification scheme and its communication modes.

## 4. Impossibilities of secure positioning

**Theorem 1:** *Suppose dishonest provers $P_1$ and $P_2$ receive quantum systems $H_1$ and $H_2$ from verifiers $V_1$ and $V_2$, which encode information $I_p$, at points $p_1(x_p^1, t_p^1)$ and $p_2(x_p^2, t_p^2)$ respectively. In general, any position-verification scheme is insecure if it allows that the position to be verified $p(x_p, t_p)$ lies in the common casual future of points $p_1(x_p^1, t_p^1)$ and $p_2(x_p^2, t_p^2)$.*

*Proof:* It can be seen from figure 2(b) that if dishonest prover $P_1$ and $P_2$ can receive encoded quantum system $H = H_1 \otimes H_2$ such that $p(x_p, t_p)$ lies in their common casual future, they can apply some superopertor $U = U_1 \otimes U_2$ locally at points $p_1(x_p^1, t_p^1)$ and $p_2(x_p^2, t_p^2)$ that transforms system $H_1 \otimes H_2$ to $U(H_1 \otimes H_2)$ instantaneously. Hence they can extract information $I'_p$ at points $q_1(x_q^1, t_q^1)$ and $q_2(x_q^2, t_q^2)$, where $t_q^1 = t_q^2 = t_p + \delta$, in their causal future through mutual classical communication. In conclusion, verifier $V_i$ cannot differentiate whether information $I'_p$ is returned from point $p(x_p, t_p)$ or $q_i(x_q^i, t_q^i)$. The position-verification schemes that cannot evade impossibility theorem 1 are shown in figure 2.

Let's write the theorem 1 in a more systematic way. Suppose verifiers $V_i$ ($i$=1,2) and $V_{\bar{i}}$ encodes information $I_p$ over quantum systems $H_i$ and $H_{\bar{i}}$ at points $(x_v^i, 0)$ and $(x_v^{\bar{i}}, 0)$ respectively. Here $\bar{i} = 2$ if $i = 1$ and vice versa. If prover receives encoded system $H_i$ and $H_{\bar{i}}$ at point $p(x_p, t_p)$, and returns information $I'_p$ to both verifiers $V_i$ and $V_{\bar{i}}$ at points $v_i(x_v^i, t_v^i)$ and $v_{\bar{i}}(x_v^{\bar{i}}, t_v^{\bar{i}})$, then theorem 1 can be restated as:

**Theorem 1:** *Suppose dishonest provers $P_i$ and $P_{\bar{i}}$ receive quantum systems $H_i$ and $H_{\bar{i}}$ from verifiers $V_i$ and $V_{\bar{i}}$, which encode information $I_p$, at points $p_i(x_p^i, t_p^i)$ and $p_{\bar{i}}(x_p^{\bar{i}}, t_p^{\bar{i}})$ respectively. In general, any position-verification scheme is insecure if it allows that all points $(x_v^i, 0)$, $p_i(x_p^i, t_p^i)$, $p(x_p, t_p)$, $q_{\bar{i}}(x_q^{\bar{i}}, t_q^{\bar{i}})$, and $v_{\bar{i}}(x_v^{\bar{i}}, t_v^{\bar{i}})$ lie on same null-like hyper-surface $N_n^i$.*

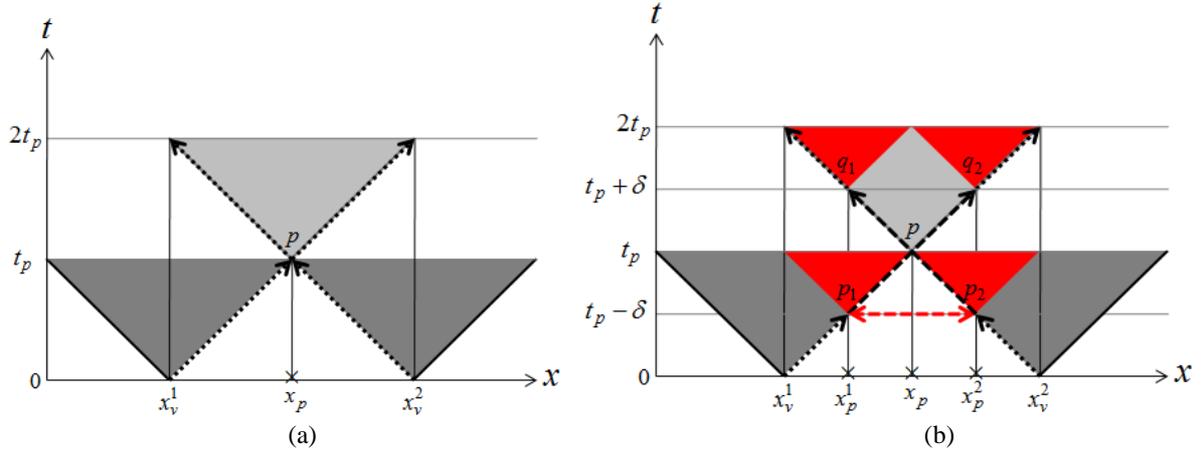

(a)                                        (b)

**Figure 2:** (a) 1+1 dimensional Minkowski space time representation of insecure quantum position-verification schemes. Dotted arrows from points $(x_v^1,0)$ and $(x_v^2,0)$ to point $p(x_p,t_p)$ represent information $I_p$ encoded over systems $H_1$ and $H_2$ while dotted arrows from point $p(x_p,t_p)$ towards its causal future represent information $I'_p$. (b) 1+1 dimensional Minkowski space time representation of successful cheating strategy from dishonest provers P1 and P2. Dashed arrows represent non-local quantum computation (red) and classical communication between $P_1$ and $P_2$.

**4.1 Existing insecure position-verification schemes:** Recently, a large number of attempts have been made to give secure positioning in relativistic quantum setup. However all these proposed schemes have been proved insecure against quantum attacks based on non-local instantaneous quantum computation [15,18]. These insecure position-verification schemes can be divided into two types based on whether provers' unitary transformation $U$ is (i) single-qubit measurements and then classical communication [12,15,18] or (ii) qubit-wise quantum operations, Bell state measurement (BSM) [31], and then classical communication [13].

In type (i), verifier $V_1$ sends information $I_p$ encoded over $H_1$ (either a pure quantum state or part of the maximally entangled pair) while verifier $V_2$ sends information about measurement basis to the prover encoded over $H_2$. The prover measures the $H_1$ in the corresponding basis obtained from $H_2$ and sends outcome $I'_p$ to both $V_1$ and $V_2$. If $I'_p = I_p$ and prover replied within allocated time, verifiers validate the position of prover as genuine. In type (ii), both $V_1$ and $V_2$ send halves $H_1$ and $H_2$ of their secret entangled pairs and classical information for corresponding unitary transformations $U_1$ and $U_2$ to the prover respectively. Prover is required to apply qubit-wise transformation $U_1 \otimes U_2 (H_1 \otimes H_2)$ and then reply his/her BSM outcome to both $V_1$ and $V_2$. If BSM result of the prover is consistent with verifiers mutual communication, then position of the prover is verified otherwise aborted.

As shown above in theorem 1, both of these approaches (i) and (ii) lead to insecure positioning since dishonest provers $P_1$ and $P_2$ can simulate their actions with the prover P. Hence the verifiers cannot be sure that whether they have received information from P at site $p(x_p,t_p)$ or from dishonest provers $P_1$ and $P_2$ at sites $p_1(x_p^1,t_p^1)$ and $p_2(x_p^2,t_p^2)$. In short, if dishonest provers $P_1$ and $P_2$ have pre-shared entanglement between them, they can break position-verification schemes of type (i) and type (ii). S. Beigi and R. Konig showed that if dishonest provers posses an exponential (in n) amount of entanglement then they can successfully attack QPV scheme of type (i) and (ii) where n qubits are communicated [32]. It has also been shown by Burrman *et al* that the minimum amount of entanglement needed to perform a successful attack on QPV schemes of type (i) and (ii) must be at least linear in the number of communicated qubits [33].

**5. Possibilities for secure positioning**
We have seen that position-verification schemes that allow verifiers, dishonest provers and the prover to apply unitary operations on same null-like hyper surface $N_n$ lead to impossibilities of secure positioning. So an obvious question would be: Can we restrict all verifiers, dishonest provers, and the prover to act unitary on same space-like hyper surface $S_T = \{t = T\}$? Interestingly, answer to this question is positive. Quantum mechanics in the form of non-locality allows verifiers and the prover to apply unitary operations simultaneously on some space-like hyper surface while causality principles bounds dishonest provers to conclude measurement outcome of their non-local instantaneous computations on same space-like hyper surface. In other words, the joint outcome of quantum mechanics and theory relativity that "any localizable superoperator is a causal superoperator" leads to the possibilities of secure positioning. Unconditionally secure positioning can only be possible if any single-round quantum position-verification scheme fulfills following theorems 2 and 3 or their unified and more restricted version, theorem 4.

**Theorem 2:** *For secure positioning, the prover must be able to extract information $I'_p$ from $H_1$ and $H_2$ at the earliest possible point $p(x_p, t_p)$ in the common causal future of both $V_1$ and $V_2$.*

**Theorem 3:** *Suppose dishonest provers $P_1$ and $P_2$ receive quantum systems $H_1$ and $H_2$ from verifiers $V_1$ and $V_2$, which encode information $I_p$, at points $p'_1(x^1_p, t'^1_p)$ and $p'_2(x^2_p, t'^2_p)$ respectively. Unconditionally secure positioning is possible if and only points $p'_1(x^1_p, t'^1_p)$ and $p'_2(x^2_p, t'^2_p)$ are causally disconnected from point $p(x_p, t_p)$, position to be verified.*

**Theorem 4:** *Unconditionally secure positioning is possible if and only if verifiers send information $I_p$ and the prover extract information $I'_p$ on a same space-like hyper surface.*

**6. Quantum scheme for secure positioning**
Here we show that causality principle can rescue single-round quantum position-verification schemes from dishonest provers having indefinite amount of pre-shared entanglement and a quantum scheme can be formulated that fulfill possibilities conditions in theorem 2 and 3 or theorem 4. Suppose verifiers $V_1$ and $V_2$ prepare quantum systems $H_{v_1} \otimes H_1$ and $H_{v_2} \otimes H_2$ at points $(x^1_v, 0)$ and $(x^2_v, 0)$ respectively. Simultaneously, both $V_1$ and $V_2$ send components systems $H_1$ and $H_2$ to P over null-like hyper surfaces connecting them to the prover without encoding information $I_p$ over quantum systems $H_1$ and $H_2$.

After time $t = t_p = (x_p - x^1_v)/c$, the prover receives component systems $H_1$ and $H_2$ and verifier $V_1$ teleports [34] information $I_p$ over EPR channel $H_{v_1} \otimes H_1$. Instantly, prover P measures his/her half $H_1$, extracts information $I'_p$ and acts as follows: He/she teleports information to verifier $V_2$ and simultaneously sends both information $I'_p$ and his/her BSM result of teleportation to both $V_1$ and $V_2$. The verifiers authenticate the position of the prover if $I'_p$ is consistent with both $I_p$ and non-local correlations generated by local measurements of verifiers and the prover on same space-like hyper surface $S_{t_p} = \{t = t_p\}$. Schematics for secure position-verification scheme that can evade quantum attacks based on non-local instantaneous computations from dishonest provers $P_1$ and $P_2$ are shown in figure 3 (a). Detailed quantum scheme for secure positioning and its security analysis can be found in our recent work [24].

It can be seen from figure 3(b) that our scheme can evade impossibility conditions in theorem 1 and 1.1 securely. Non-local instantaneous computations of dishonest provers $P_1$ and $P_2$ at points $p_1(x_p^1, t_p^1)$ and $p_2(x_p^2, t_p^2)$ cannot help dishonest provers to extract information $I'_p$ while being on the null-like hyper surface $N_n^i$ connecting verifier $V_i$ with the prover P at point $p(x_p, t_p)$.

Similarly, our scheme fulfills possibility condition stated in theorem 4. The point where $V_1$ teleports information $I_p$ to P, the point where P process systems $H_1$ and $H_2$ through unitary transformations, and the point where $V_2$ receives teleported information $I'_p$ from P, all lie on same space-like hyper surface $S_{t_p} = \{t = t_p\}$. As a result, any localizable operator $U = U_1 \otimes U_1$, which can transform system $H_1 \otimes H_2$ to $U(H_1 \otimes H_2)$ at points $p'_1(x_p^1, t_p'^1)$ and $p'_2(x_p^2, t_p'^2)$ lying on same hyper surface $S_{t_p} = \{t = t_p\}$, does not allow dishonest provers $P_1$ and $P_2$ to reply information $I'_p$ to both $V_1$ and $V_2$ in time. That is, it can be seen from figure 3(b) that in quantum position-verification scheme built on some space-like hyper surface, dishonest provers can only extract information $I'_p$ at points $q_1(x_p^2, 2t_p)$ and $q_2(x_p^2, 2t_p)$ and hence can reply correct information to both $V_1$ and $V_2$ not before $2t_p + \delta$.

Although, possibility condition stated in theorem 2 is necessary for any quantum position-verification scheme, however, it has special importance while considering tasks on some space-like hyper surface. In our scheme for secure positioning, if $V_1$ and $V_2$ are not on same space-like hyper surface, dishonest provers $P_1$ and $P_2$ can simulate their actions with the prover P through semi-localizable operations. For example, in our teleportation-based QPV scheme, dishonest provers $P_1$ and $P_2$ can manage to have a whole system $H_1 \otimes H_2$ at site of $P_1$(or $P_2$) who can then send information $I'_p$ to $P_2$(or $P_1$) through semi-localizable operations.

Finally, teleportation is the necessary quantum communication mode to insure possibility conditions stated in theorem 3 or theorem 4 that guarantees security against localizable operations of $P_1$ and $P_2$. Since provers' site can have some finite radius δ, possibility condition stated in theorem 3 can be insured without restricting verifier $V_1$ on same space-like hyper surface $S_{t_p} = \{t = t_p\}$. That is, secure positioning is possible even if $V_1$ teleports information $I_p$ at time t where $(t_p - \delta)/c < t \leq t_p$.

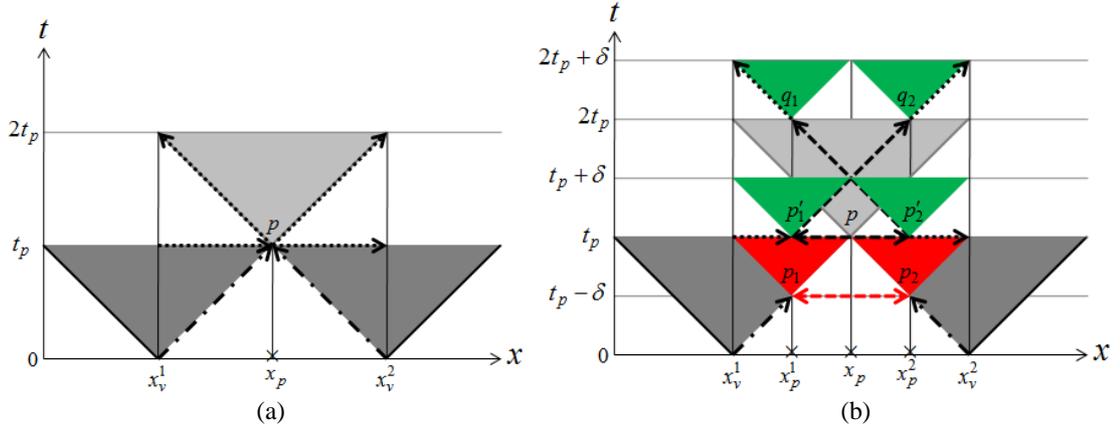

(a) (b)

**Figure 2:** (a) 1+1 dimensional Minkowski space time representation of secure quantum position-verification schemes. Dotted-dashed arrows from points $(x_v^1,0)$ and $(x_v^2,0)$ to point $p(x_p,t_p)$ systems $H_1$ and $H_2$ while dotted arrows (horizontal) represent information $I_p$ teleported from verifier $V_1$ to prover at point $p(x_p,t_p)$ and information $I'_p$ from point $p(x_p,t_p)$ toward its causal future. (b) 1+1 dimensional Minkowski space time representation of unsuccessful cheating strategy from dishonest provers P1 and P2. Dashed arrows represent non-local quantum computation (horizontal) and classical communication between $P_1$ and $P_2$.

## 7. Discussion

We showed that an enigma of secure positioning in both classical and quantum information is solvable even if sender and receiver have no pre-shared quantum/classical data while dishonest provers have efficient quantum technology and unlimited computational powers. We showed that previously existing no-go theorem for secure positioning is the upshot of not utilizing full powers of relativistic quantum information theory and prove that secure positioning is possible if causal structure of Minkowski space time and quantum non-locality is utilized properly.

By using geometric structure of Minkowski space time, we analyzed the impossibility for secure positioning on the basis of quantum communication modes and causality principle. We also showed that secure positioning is possible and comprehensively outlined possibility conditions that have to be followed by every secure position verification scheme.

The outcome can be summarized as follows. Single-round quantum position-verification schemes based on qubit-wise unitary transformations on some null-like hyper surface are proved to be insecure. In the existing quantum communication modes, only option for achieving unconditionally secure positioning is teleportation. Teleportation is necessary and sufficient mode for formulating secure positioning where bother verifiers and the prover apply unitary transformations on same space-like hyper surface. Here, causality principle from the theory of relativity rescues the teleportation-based quantum position verification constructed over some space-like hyper surface. If the verifier teleports and the prover receive information at some space-like hyper surface, then any quantum operation that allows space like-separated dishonest provers $P_1$ and $P_2$ to communicate and extract encoded information is not *causal* and hence not physically implementable.

This possibility of secure positioning would have applications in long distance communication and other quantum tasks. For example, if prover P is considered to be a quantum repeater, then the parties $V_1$ and $V_2$ can use its secure positioning for long distance communications. Instead of sending message directly, $V_1$ sends encrypted message to P who then passes the message to $V_2$. If $V_1$ and $V_2$ agree on the position verification of P, then $V_1$ can communicate and reveal message to $V_2$. Moreover, some quantum tasks such as quantum multiplayer games strictly require that certain players (dishonest in general) are not allowed to communicate with each other. Non-relativistic quantum information theory does not guarantee this security requirement for multiplayer games. However, relativistic quantum information and hence secure positioning can ensure that each player is performing local unitary operations from some specified position and is restricted from communication with other players through causality principle.

Finally, we assumed that the verifiers and provers, honest or dishonest, lie in same inertial frame. The assumption is quite acceptable for many practical applications such as communication between military bases, communication between a customer and bank in nearby vicinity, automatic toll collection at some specified locations etc.

However, to make it more general from moving provers on earth to earth-satellite communications, we need to generalize this procedure where Lorentz invariance is necessary ingredient. Our proposed possibility conditions for secure positioning are in fact an important step towards this generalization. These conditions guarantee secure positioning by bounding verifiers and the prover to transmit and receive on same space-like hyper surface in specific order. This bound on ordering of quantum unitary operation is an important step for Lorentz invariance.